\documentclass[twocolumn,prl,showpacs,amsmath,floatfix]{revtex4}
\usepackage{graphicx}

\begin{document}

\title{Emergence of Complex Spatio-Temporal Behavior in Nonlinear Field
Theories}

\author{Marcelo Gleiser}
\email{mgleiser@dartmouth.edu}
\affiliation{Department of Physics and Astronomy, Dartmouth College, Hanover,
NH 03755, USA}

\author{Rafael C. Howell}
\email{rhowell@lanl.gov}
\affiliation{Materials Science and Technology Division, Los Alamos National
Laboratory, Los Alamos, NM 87545, USA}

\date{\today}

\begin{abstract}
We investigate the emergence of time-dependent nonperturbative configurations
during the evolution of nonlinear scalar field models with symmetric and
asymmetric double-well potentials. Complex spatio-temporal behavior emerges as
the system seeks to establish equipartition after a fast quench. We show that
fast quenches may dramatically modify the decay rate of metastable states in
first order phase transitions. We briefly suggest possible applications in
condensed matter systems and early universe cosmology.

\centering LA-UR 06-2478
\end{abstract}



\maketitle

\section{Introduction}

The emergence of complex patterns is one of the most distinct signatures of
nonlinear interactions in natural systems. Since Einstein's pioneering work on
Brownian motion \cite{Einstein05}, it became clear that much can be
accomplished by modeling the interactions of a system with its environment
through the action of random and viscous forces. During most of the twentieth
century, studies were mainly restricted to investigating the motion of a point
particle in nonlinear potentials \cite{Brownian}. With the advent of fast
computers, modeling of stochastic evolution added spatial dimensions, allowing
for the quantitative study of spatio-temporal complex behavior. Up to about ten
years ago, most of the work concentrated in hydrodynamical and soft
condensed-matter systems \cite{Walgraef}. Recently, developments in high energy
physics and cosmology have opened the interesting possibility that complex
spatio-temporal behavior may also play a role in relativistic field theories,
in particular during the early stages of cosmological evolution
\cite{cosmology} and may even be observed in high-energy colliders
\cite{doscil}.

Here we will briefly review some of the work done during the past few years
which focused on understanding the effects of fast quenches on nonlinear scalar
field theories. The quenches model both temperature quenches in the context of
fast cosmological expansion (in particular at scales close to the GUT scale
$\sim 10^{16}$ GeV) or the cooling of fireballs during high energy collisions
such as those currently being investigated at RHIC and soon at LHC. The quench
may also represent a pressure quench, common in condensed matter physics or,
more generally, the appearance of a low-energy effective interaction that
modifies the effective potential of the long-wavelength modes of the field or
order parameter describing the system's evolution.

\section{The Model}
Consider a (2+1)-dimensional real scalar field (or scalar order parameter)
$\phi({\bf x},t)$ evolving under the influence of a potential $V(\phi)$. The
continuum Hamiltonian is conserved and the total energy of a given field
configuration $\phi({\bf x},t)$ is, 
\begin{equation} H[\phi]=
\int d\,^2x\left[\frac{1}{2}(\partial_{t}\phi)^2+\frac{1}{2}(\nabla\phi)^2
+V(\phi)\right],
\label{H} 
\end{equation} 
where
$V(\phi)=\frac{m^2}{2}\phi^2-\frac{\alpha}{3}\phi^3+\frac{\lambda}{8}\phi^4$ 
is the potential energy density. The parameters $m$, $\alpha$, and $\lambda$
are positive definite and temperature independent.  It is helpful to introduce
the dimensionless variables $\phi'=\phi\sqrt\lambda/m$, $x'=xm$, $t'=tm$, and
$\alpha'=\alpha/(m\sqrt\lambda)$ (We will henceforth drop the primes).  Prior
to the quench, $\alpha=0$ and the potential is an anharmonic single well
symmetric about $\phi=0$. The field is in thermal equilibrium with a
temperature $T$. At the temperatures considered, the fluctuations of the field
are well approximated by a Gaussian distribution, with
$\langle\phi^2\rangle=aT$ ($a=0.51$ and can be computed numerically). As such,
within the context of the Hartree approximation \cite{aarts}, the momentum and
field modes in $k$-space can be obtained from a harmonic effective potential,
and satisfy $\langle|{\bar\pi}(k)|^2\rangle=T$ and
$\langle |\bar{\phi}({\bf k})|^2\rangle=\frac{T}{k^2+m_H^2}$, respectively. 
The Hartree mass $m_H^2=1+\frac{3}{2}\langle\phi^2\rangle$ depends on the 
magnitude of the fluctuations (and thus $T$). Within the Hartree approximation
we can write the effective potential as
\begin{widetext}
\begin{equation}
V_{\rm eff}\left(\phi_{\rm ave},m_{\rm H}^2\right)=
\left[1-m_{\rm H}^2(t)\right]\phi_{\rm ave}
+\frac{1}{2}\,m_{\rm H}^2(t)\,\phi_{\rm ave}^2
-\frac{\alpha}{3}\,\phi_{\rm ave}^3+\frac{1}{8}\,\phi_{\rm ave}^4\,.
\label{Veff}
\end{equation}
\end{widetext}
Hereafter we will refer to a particular system by its initial temperature. All
results are ensemble averages over 100 simulations.

If $\alpha\neq 0$, the ${\cal Z}_2$ symmetry is explicitly broken. When
$\alpha=1.5\equiv\alpha_c$, the potential is a symmetric double-well (SDW),
with two degenerate minima. This is the first case we consider.

\section{Quenching Into Symmetric Double Wells: Emergence of Spatio-Temporal
Order}

\begin{figure}
\includegraphics[width=3in]{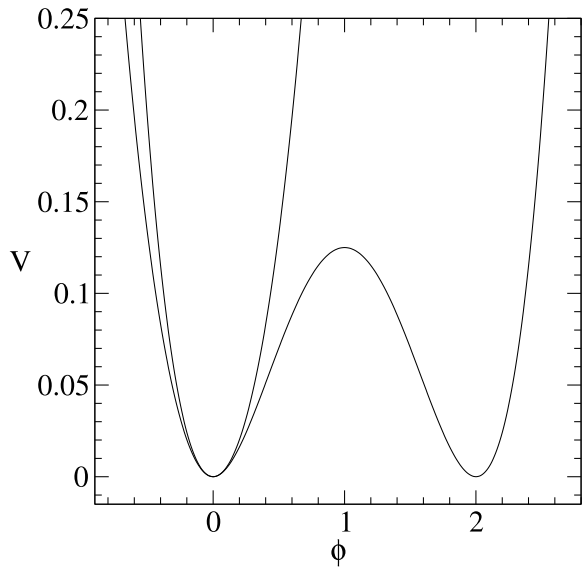}
\caption{Schematic picture showing change in potential $V(\phi)$ from
single-well to symmetric double-well after quench.}
\label{sympot}
\end{figure}

At $\alpha=\alpha_c=1.5$, the quench amounts to switching from a single to
a double well with the field initially localized at $\phi=0$. In Fig.
\ref{sympot} we indicate this schematically. 

As shown in Ref. \cite{Gleiser-Howell}, the quench sets oscillations in the
field's zero mode, $\phi_{\rm ave}\equiv\frac{1}{V}\int\phi\,dV$, where $V$ is
the volume (or area in 2d). The amplitude of these oscillations is controlled
by the temperature of the initial Gaussian distribution, as explained above. In
fact, temperature here is simply a convenient way to set an initial Gaussian
distribution in momentum space. We did this using a Langevin equation with
white noise. One could state that the field is at $T=0$ but initially set with
a Gaussian distribution in momentum space with a certain width. This width is a
measure of the initial ``temperature'' of the system.

At early times small fluctuations satisfy a Mathieu equation in $k$-space
\begin{equation}
\ddot{\delta\phi}=
-\left[ k^2+V_{\rm eff}''\left[\phi_{\rm ave}(t)\right]\right]\delta\phi,
\label{flucteq}
\end{equation}
and, depending on the wave number and parametric oscillations of
$\phi_{\rm ave}(t)$, can undergo exponential amplification ($\sim e^{\eta t}$).
For $T\leq 0.13$, no modes are ever amplified. As the temperature is increased,
so is the amplitude and period of oscillation in $\phi_{\rm ave}$, gradually
causing the band $0<k<0.48$ to resonate and grow. Furthermore, for large enough
temperatures ($T>0.13$) large-amplitude fluctuations about the zero mode probe
into unstable regions where $V_{\rm eff}''<0$, which also promote their growth. 
Note that this is very distinct from spinodal decomposition, where competing
domains of the two phases coarsen \cite{gunton}. Instead, for the values of $T$
and $\alpha$ considered, $\phi_{\rm ave}$ continues to oscillate about the
$\phi=0$ minimum.

\begin{figure}
\includegraphics[width=3in]{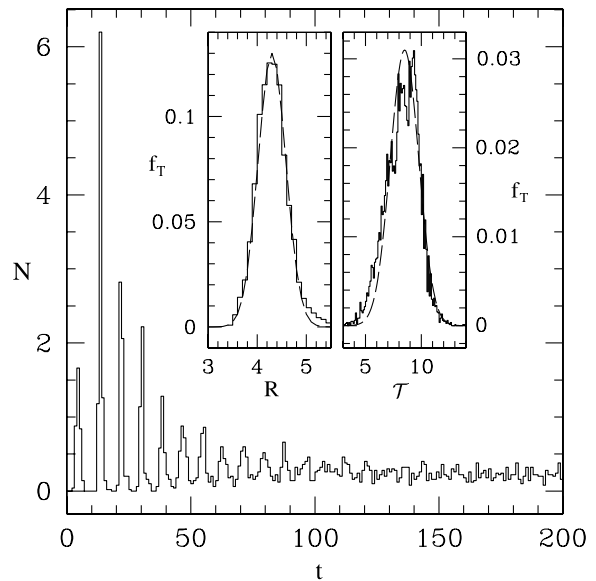}
\caption{Number of oscillons nucleated between $t$ and $t+\delta t$ at $T=0.22$
and $\delta t=1$. The global emergence is evident early in the simulations.
Inset: probability distribution of radii and periods of oscillations of
individual oscillons.}
\label{oscillons_count}
\end{figure}

As a result of the energy transfer modeled by parametric amplification,
oscillons are nucleated initially in phase. But what are oscillons? They are
the higher dimensional equivalent of kink-antikink breathers, familiar of 1d 
nonlinear dynamics \cite{campbell}. Extensive work has been done on oscillons
and their properties and the reader can consult the relevant literature listed
in Ref. \cite{oscillons}. Here, it is enough to mention that oscillons are
long-lived, time-dependent, localized field configurations which express local
ordering of momentum modes. What was also observed in Ref.
\cite{Gleiser-Howell} is that after the quench oscillons emerge in synchrony,
exhibiting both spatial and time ordering. In Figure \ref{oscillons_count} we
illustrate this phenomenon.

\begin{figure}
\includegraphics[width=3in]{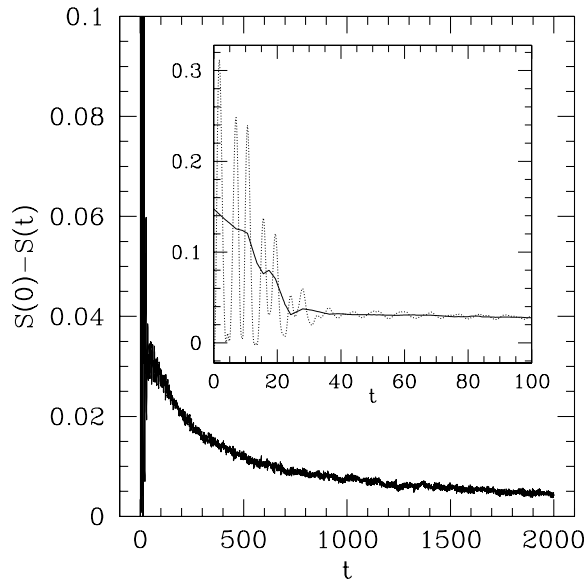}
\caption{The change of $\Pi(t)$ from the initial state for closed systems at
$T=0.22$. The exponential approach to equilibrium is clear at late times. The
inset illustrates the role of oscillons as a bottleneck to equipartition.}
\label{entropy}
\end{figure}

Finally, we introduce a measure of the partitioning of the kinetic energy
$\Pi(t)$, which we use to describe the nonequilibrium evolution of the system:
\begin{equation} 
\Pi(t)=-\int d^2 k~p({\bf k},t) \ln p({\bf k},t),
\end{equation} 
where $p({\bf k},t)=K({\bf k},t)/\int d^2 k K({\bf k},t)$, and $K({\bf k},t)$
is the kinetic energy of the $k$-th mode. $\Pi(t)$ attains its maximum
($\Pi_{\rm max}=\ln(N)$ on a lattice with $N$ degrees of freedom) when
equipartition is satisfied. This occurs both at the initial thermalization
$(t=0)$ and final equilibrium states, since in this case all modes carry the
same fractional  kinetic energy. In Fig. \ref{entropy} we show the change of
$\Pi(t)$ from the initial state,  $\Pi(t=0)-\Pi(t)$, for the closed system at
$T=0.22$. At late times ($t>150$), we have found that the system equilibrates
exponentially in a time-scale $\tau\simeq 500$. At early times, the
localization of energy at lower ${\bf k}$-modes, corresponding to the global
emergence of oscillons, prolongs this approach to equipartition. The inset of
Fig. \ref{entropy} shows the large variations in $\Pi(t)$ (dotted  line) that
arise due to the synchronous oscillations in the kinetic energy of these
configurations. Also shown (solid line) is the average between successive
peaks of $\Pi(t)$, with a plateau at approximately $20<t<70$ that coincides
with the maximum oscillon presence in the system. Thus, oscillon
configurations serve as early bottlenecks to equipartition, temporarily
suppressing the  diffusion of energy from low ($0<|{\bf k}|\leq 0.8$) to higher
modes.

\section{Quenching Into Asymmetric Double Wells: Resonant Nucleation}

\begin{figure}
\includegraphics[width=3in]{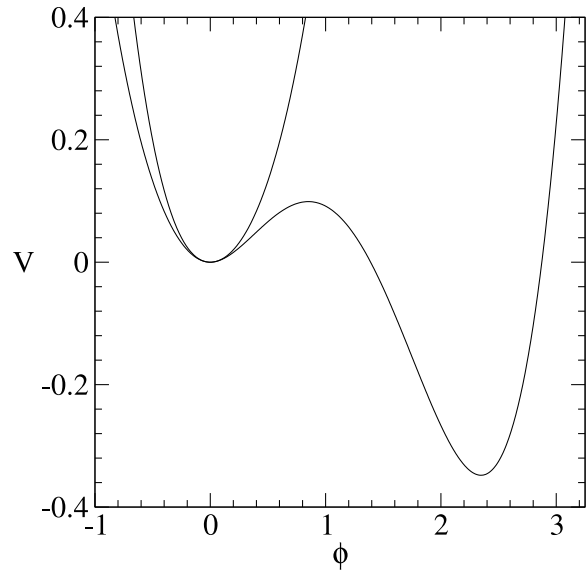}
\caption{Schematic picture showing change in potential $V(\phi)$ from
single-well to asymmetric double-well after quench.}
\label{asympot}
\end{figure}

For $\alpha>\alpha_c=1.5$ the potential is asymmetric with the minimum at
$\phi=0$ becoming metastable. We proceed as before by quenching the system from
a single well, as illustrated in Fig. \ref{asympot}.

As before, oscillons will once again be nucleated. However, the situation now
changes dramatically. Due to the asymmetry, the system will decay into the
global minimum at $\phi_+$. We have observed that this decay may occur in three
possible ways depending on the initial temperature $T$ and the value of
$\alpha$ \cite{Gleiser-Howell2}: i) the transition to the global minimum
happens very fast in what can be called a ``cross-over'' transition; ii) the
transition occurs as a single oscillon becomes unstable and grows into a
critical bubble. As is well know from the theory of first order phase
transitions \cite{gunton}, once a critical nucleus forms it will grow to
complete the transition; iii) two or more oscillons coalesce to become a
critical bubble that then grows to complete the transition. 

In order to simplify the analysis, we fixed the temperature to be $T\leq 0.22$.
From the Hartree potential of Eq. \ref{Veff}, one can see that for large
temperatures the potential becomes a single well again. For $T\leq 0.13$ no
oscillons are nucleated after the quench. In this case, we expect that the
usual metastable decay rate based on the theory of homogeneous nucleation (HN)
will apply, becoming more accurate for smaller $T$ \cite{gunton, Coleman}. The
decay rate per unit volume obtained from HN theory is controlled by the
Arrhenius exponential suppression,
$\Gamma(T,\alpha)\simeq T^{(d+1)}\exp[-E_b(T,\alpha)/T]$, where $E_b$ is the
energy of the critical bubble or nucleus and $d$ is the number of spatial
dimensions. [We use units where $c=k_B=\hbar=1$.] The typical time-scale
for the decay in a volume $V$ is then,
$\tau_{\rm HN}\simeq (V\Gamma)^{-1}\sim T^{-1}\exp[E_b(T,\alpha)/T]$.

\begin{figure}
\includegraphics[width=3in]{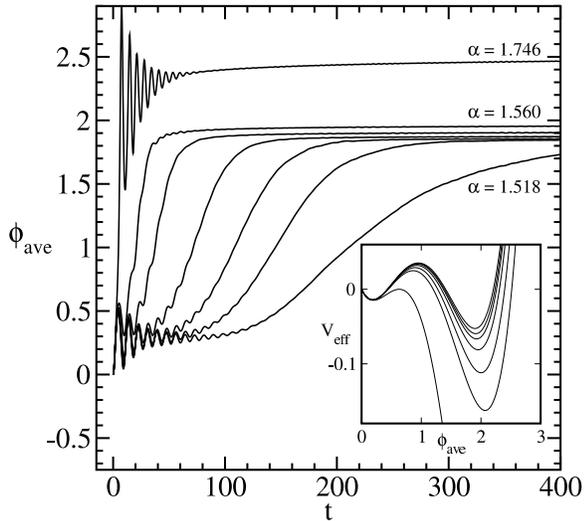}
\caption{The evolution of the order parameter $\phi_{\rm ave}(t)$ at $T=0.22$
for several values of the asymmetry. From left to right,
$\alpha= 1.746, 1.56, 1.542, 1.53, 1.524, 1.521, 1.518$. The inset shows
$V_{\rm eff}$ for the same values.}
\label{decay-phi}
\end{figure}

In Fig. \ref{decay-phi} we show the evolution of the order parameter
$\phi_{\rm ave}(t)$ as a function of time for several values of asymmetry,
$1.518\leq\alpha\leq 1.746$, for $T=0.22$. Not surprisingly, as
$\alpha\rightarrow\alpha_c=1.5$, the field remains longer in the metastable
state, since the nucleation energy barrier $E_b\rightarrow\infty$ at $\alpha_c$.
However, a quick glance at the time axis shows the fast decay time-scale, of
order $10^{1-2}$. For comparison, for $1.518\leq\alpha\leq 1.56$, HN would
predict nucleation time-scales of order
$\sim 10^{28}\geq\tau_{\rm HN}\sim\exp[E_b/T]\geq 10^{12}$ (in dimensionless
units). [The related nucleation barriers with the effective potential are
$E_b(\alpha=1.518)=14.10$ and $E_b(\alpha=1.56)=5.74$.] For small asymmetries
$\phi_{\rm ave}(t)$ displays similar oscillatory behavior to the SDW case
before transitioning to the global minimum. As $\alpha$ is increased the number
of oscillations decreases. For large asymmetries, $\alpha\geq 1.746$, the
entire field crosses over to the global minimum without any nucleation event,
resulting in oscillations about the global minimum. This is the situation
described in case i) above.

\begin{figure}
\includegraphics[width=3in]{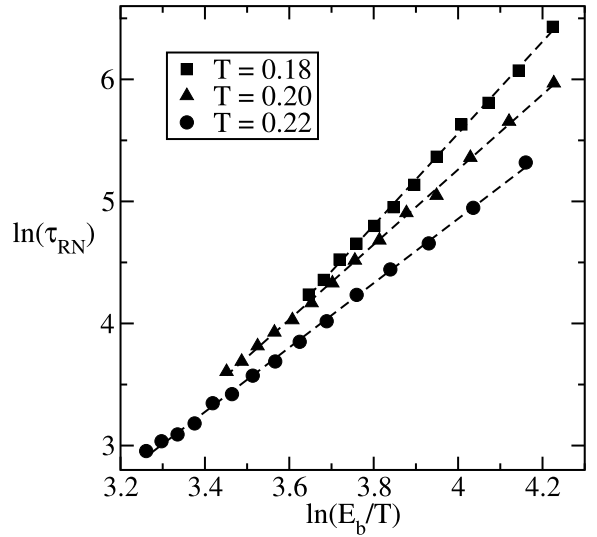}
\caption{Decay time-scale $\tau_{\rm RN}$ as a function of critical nucleation
effective free-energy barrier $E_b/T$ at $T=0.18$, $0.20$, and $T=0.22$. The
best fits (dashed lines) are power-laws with exponents $B\simeq 3.762$, $3.074$,
and $2.637$, respectively.}
\label{powerlaw}
\end{figure}

In Fig. \ref{powerlaw} we show the ensemble-averaged nucleation time-scales for
resonant nucleation, $\tau_{\rm RN}$, as a function of the nucleation barrier
(computed with Eq. \ref{Veff}), $E_b/T$, for the temperatures
$T=0.18$, $0.20$, and $0.22$. [For temperatures above $T=0.26$ we are in the
vicinity of the critical point in which no barrier exists.] The nucleation time
was measured when $\phi_{\rm ave}$ crosses the maximum of $V_{\rm eff}$. The
best fit is a power law:
\begin{equation}
\tau_{RN}\propto (E_b/T)^B, 
\end{equation}
with $B=3.762\pm 0.016$ for $T=0.18$, $B=3.074\pm 0.015$ for $T=0.20$, and
$B=2.637\pm 0.018$ for $T=0.22$. This simple power law holds for the same range
of temperatures where we have observed the synchronous emergence of oscillons.
It is not surprising that the exponent $B$ increases with decreasing $T$, since
the synchronous emergence of oscillons becomes less pronounced and eventually
vanishes. In these cases we should expect a smooth transition into the
exponential time-scales of HN. 

We conclude that {\it fast quenching can dramatically affect the nucleation
time-scale of first order phase transitions}. In other words, HN fails for fast
quenches.

\begin{figure}
\includegraphics[width=2in]{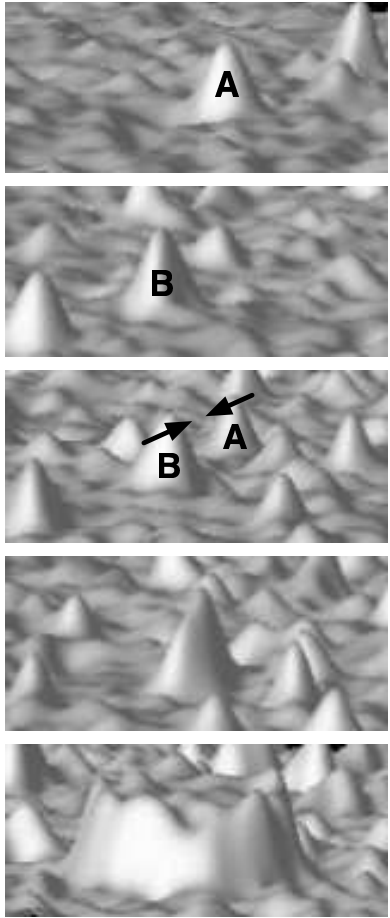}
\caption{Two oscillons coalesce to form a critical bubble. First two frames
from top show oscillons A and B. Third and fourth frames shows A and B
coalescing into a critical bubble. Final frame shows growth of bubble expanding
into metastable state.}
\label{oscil_coal}
\end{figure}

Here we propose the mechanism by which this fast decay occurs: for nearly
degenerate potentials, $\alpha_c<\alpha\leq\alpha_{\rm I}$, the critical
nucleus has a much larger radius than a typical oscillon; it will appear as two
or more oscillons coalesce. We call this Region I, defined for
$R_b\geq 2R_{\rm osc}$, where $R_{\rm osc}$ is the minimum oscillon radius
computed from Ref. \cite{doscil}. Figure \ref{oscil_coal} illustrates this
mechanism. Two oscillons, labeled A and B, join to become a critical nucleus.
[The interested reader can see simulation movies at
http://www.dartmouth.edu/$\sim$cosmos/oscillons.]

\begin{figure}
\includegraphics[width=3in]{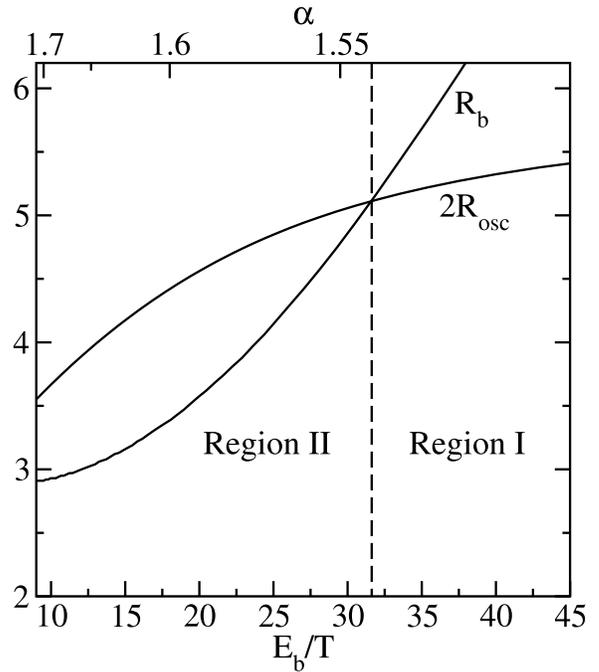}
\caption{Radius of critical bubble ($R_b$) and twice the minimum oscillon
radius ($2R_{\rm osc}$) as a function of its energy barrier and related values
of $\alpha$ at $T=0.22$. For $\alpha \geq 1.547$ one cannot easily distinguish
between an oscillon and a critical bubble.}
\label{radius}
\end{figure}

As $\alpha$ is increased further, the radius of the critical nucleus decreases,
approaching that of an oscillon. In this case, a single oscillon grows
unstable to become the critical nucleus promoting the fast decay of the
metastable state: there is no coalescence. We call this Region II,
$\alpha_{\rm I}<\alpha\leq\alpha_{\rm II}$, $R_b<2R_{\rm osc}$. This explains
the small number of oscillations on $\phi_{\rm ave}(t)$ as $\alpha$ is
increased  [cf. Fig. \ref{decay-phi}]. To corroborate our argument, in
Fig. \ref{radius} we contrast the critical nucleation radius with that of
oscillons as obtained in Ref. \cite{doscil}, for different values of effective
energy barrier and related values of $\alpha$ at $T=0.22$. The critical nucleus
radius $R_b$ is equal to $2R_{\rm osc}$ for $\alpha=1.547$. This defines the
boundary between Regions I and II: for $\alpha\geq\alpha_{\rm I}$ a single
oscillon may grow into a critical bubble. Finally, for
$\alpha\geq\alpha_{\rm II}=1.746$ the field crosses over to the global minimum
without any nucleation event.

An obvious extension of the present work is the investigation of ``resonant
nucleation'' in 3d. Preliminary results indicate that the power law behavior
persists with $B\sim 1.5$ for the relevant range of temperatures for oscillon
coalescence. An application in cosmology has recently been suggested in the
context of a two-field model of inflation \cite{Gleiser}. In general, RN will
occur whenever the effective potential changes faster than the typical
relaxation rate of the longest wavelength of the order parameter. These results
could be extended to systems in the Ising universality class, in particular to
ferromagnetic transitions, where the quench could be induced by a homogeneous
external magnetic field. These topics are presently under investigation.

\begin{acknowledgments}
One of us (MG) would like to thank the organizers, in particular Jean-Michel
Alimi for being a wonderfully gracious host.
\end{acknowledgments}

\end{document}